# Magnetoresistance due to Broken $C_4$ Symmetry in Cubic B20 Chiral Magnets

S. X. Huang[1*,#], Fei Chen[1,3], Jian Kang[2], Jiadong Zang[1*], G. J. Shu[4], F. C. Chou[4], and C. L. Chien[1*]

[1]*Department of Physics and Astronomy, Johns Hopkins University, Baltimore, MD 21218, USA*

[2]*School of Physics and Astronomy, University of Minnesota, Minneapolis, MN 55455, USA*

[3]*Department of physics, Nanjing University, Nanjing 210093, China*

[4]*Center for Condensed Matter Sciences, National Taiwan University, Taipei 10617, Taiwan*

Abstract:

The B20 chiral magnets with broken inversion symmetry and $C_4$ rotation symmetry have attracted much attention. The broken inversion symmetry leads to the Dzyaloshinskii-Moriya that gives rise to the helical and Skyrmion states. We report the unusual magnetoresistance (MR) of B20 chiral magnet $Fe_{0.85}Co_{0.15}Si$ that directly reveals the broken $C_4$ rotation symmetry. We present a microscopic theory, a minimal theory with two spin-orbit terms, that satisfies all the symmetry requirements and accounts for the transport experiments.

Broken symmetry is a fundamental concept prevailing in many branches of physics. The physical properties of a crystalline solid are intimately linked to its symmetry. Any broken symmetry in the solid is likewise consequential. This is evident in the cubic non-centrosymmetric B20 chiral magnets (MnSi, FeGe, $Fe_{1-x}Co_xSi$ etc.), which have attracted much attention because they harbor the helical and the Skyrmion states [1-4] due to the broken inversion symmetry. The B20 magnets, like the well-known ferromagnets of Fe and Ni, also have a cubic Bravais lattice, but without the inversion symmetry and the 4-fold ($C_4$) rotation symmetry. The broken inversion symmetry leads to the Dzyaloshinskii-Moriya (D-M) interaction [5, 6], in addition to the Heisenberg exchange interaction. The D-M interaction with energy $D\boldsymbol{M}\cdot(\nabla\times\boldsymbol{M})$, where $D$ is the Dzyaloshinskii constant and $\boldsymbol{M}$ is the magnetization, favors perpendicular spin alignment, as opposed to the collinear spin alignment demanded by the Heisenberg interaction with energy $A(\nabla M)^2$, where $A$ is the exchange stiffness constant. The competition between the Heisenberg and the D-M interactions leads to a spin helix ground state [7, 8] with zero net magnetization, instead of the usual ferromagnetic ground state. More interestingly, at temperatures close to but below the Curie temperature ($T_C$), and under a magnetic field, an exotic magnetic Skyrmion state [2, 3] emerges with a non-trivial topology. The spin helix and Skyrmion state in the B20 magnets, first revealed by neutron diffraction and Lorentz transmission electron microscopy [1, 3, 4], have captured much attention for its intriguing physics such as the Skyrmion lattice, the topological Hall effect [9-11], the emergent electromagnetic field, and unique prospects for applications [12-14]. However, there has been no observation of the consequential properties due to the broken $C_4$ rotation symmetry in the B20 magnets. We report in this work the observation of the unusual magnetoresistance (MR) in (Fe-Co)Si, a prototype B20 magnet, with characteristics different from those of their counterparts in cubic centrosymmetric ferromagnets. Equally important, the experimental observation also places constraints on the Hamiltonian for describing the physics of B20 materials. Our Hamiltonian with two spin-orbit terms fulfills all the symmetry requirements can account for the experimental results.

The B20 structure (e.g. FeSi) has a cubic unit cell with lattice constants $a = b = c$ (Fig. 1a). Each unit cell consists of four Fe atoms and four Si atoms with coordinates $(u,u,u)$, $(0.5+u,0.5-u,-u)$, $(-u, 0.5+u, 0.5-u)$, $(0.5-u,-u,0.5+u)$, where $u$(Fe)=0.1358 and $u$(Si)=0.844 [15]. This structure can also be constructed from a face-centered cubic (FCC) lattice, which can be decomposed into four equivalent interpenetrating simple cubic (SC) sublattices of lattice constant $a$. One obtains the B20 structure by placing the Fe-Si units with four different orientations (parallel to the 4 body diagonal directions) onto the four SC sublattices. Comparing with the FCC lattice with the full octahedral symmetry, the inversion and $C_4$ rotation symmetries are broken in the cubic B20 compounds. While the broken inversion symmetry is obvious in the B20 unit cell (Fig. 1a), the broken $C_4$ symmetry is more subtle and its consequences on the magnetic and transport properties were previously unknown.

Cubic B20 FeSi is an insulator whereas CoSi is a paramagnetic metal. For the composition range of $0.05 < x < 0.8$, $Fe_{1-x}Co_xSi$ is a conducting B20 magnet with a maximum $T_C$ of about 50K. In this work, we report the observation of a unique MR as a result of the broken $C_4$ symmetry in $Fe_{0.85}Co_{0.15}Si$ with characteristics different from all other MRs, including the well-known anisotropic MR (AMR) in polycrystalline or single crystalline centrosymmetric cubic ferromagnets such as Ni.

We used a $Fe_{0.85}Co_{0.15}Si$ single crystal sample (about 0.45mm × 0.45mm × 5 mm) with square cross section with edges parallel to the $x$[001], $y$[010], and $z$[001] directions for the magnetic and transport measurements (Fig. 1a). The sample has $T_C \approx 23$ K as revealed by the AC susceptibility measurement (inset of Fig. 1b). The hallmarks of common ferromagnets are the *M-H* curves showing distinct magnetic anisotropy, magnetic hysteresis and finite remanence at zero field. In contrast, the *M-H* curves of $Fe_{0.85}Co_{0.15}Si$ show no hysteresis and zero remanence due to the helical ground state with zero net magnetization. The *M-H* curve is quasi linear in field until the saturation field ($H_S$) [16] due to the formation of the conical phase under a magnetic field. The value of $H_S$ is the sum of the demagnetization field due to shape anisotropy and $H_D=D^2M/2A$, the critical field of the conical phase transforming into ferromagnetic alignment, where $D$ is the Dzyaloshinskii-Moriya coefficient and $A$ is the exchange stiffness mentioned above. Specifically, the *M-H* curves are the *same* for fields along [010] and [001] directions, as shown in Fig. 1b even though the $C_4$ symmetry is broken. Aside from a small difference in the demagnetization field, the same result has also been observed for the [100] direction. Thus, the B20 magnet $Fe_{0.85}Co_{0.15}Si$ shows complete magnetic *isotropy* with the same characteristics with fields along the three crystalline axes of [100], [010], and [001]. The reason is that the ferromagnetic state is a long-range order with a characteristic length scale much larger than that of the lattice constant. Thus, the broken $C_4$ symmetry in the unit cells of B20 is *not* revealed in the magnetic properties.

In contrast, broken $C_4$ symmetry may be seen in the transport measurements since the momenta of the conduction electrons, with shorter characteristic length scales comparable to the lattice constant, can probe inside the unit cells. In the transport measurements, we apply a current along the $x$([100]) direction and measure the MR as a function of field along the $x$([100]), $y$([010]), and $z$([001]) directions. We also scan a magnetic field of 2 kOe (larger than $H_S \approx 1.2$ kOe) in the *xy, yz,* and *zx* crystal planes to reveal the angular dependences of the MR. Below $H_S$, the field dependence of MR due to spin texture correlates with that of the *M-H* curve. Above $H_S$, with the magnetic moments already aligned, the resistance is usually unchanged. Peculiar to the B20 $Fe_{1-x}Co_xSi$ magnets, the resistance continues to increase linearly with field unabated even at a very large magnetic field. This intriguing positive linear MR, also previously observed, may have a complex origin including quantum interference effects [17] and the Zeeman shift of the exchange-split bands [18]. As shown in the inset of Fig. 1c, this linear positive MR is independent of the field direction, weakly dependent on temperature (at $T<T_C$) and exists even above $T_C$ [17, 18], suggesting a non-ferromagnetic origin. Despite of the same *M-H* curves, the field dependence of MR is completely *different* for field along the *y*([010]), and *z*([001]) directions.

To recognize the unusual MR results in the B20 magnets, it is useful to first describe the well-known AMR behavior in centrosymmetric ferromagnets such as Ni and permalloy (Py) [19, 20]. In polycrystalline ferromagnetic metals, the resistivity $\rho$ depends only on the angle ($\varphi$) between the directions of the electric current (***I***) direction (defined as *x* axis) and magnetization (***M***) with an axial symmetry of

$$\rho = \rho_\perp + (\rho_\parallel - \rho_\perp)\cos^2\varphi, \qquad (1)$$

where $\rho_\parallel$ and $\rho_\perp$ are the resistivities with ***M*** parallel and perpendicular to ***I*** respectively. Using a sufficiently large ***H*** to align ***M***, both the *xy* and the *zx* scans show the same AMR magnitude of $(\rho_\parallel - \rho_\perp)$ and a $\cos^2\varphi$ angular dependence, which is a general angular dependence of anisotropic conduction. Most notably, the *yz* scan, with ***M*** perpendicular to ***I***, shows *no* variation at all (Fig. 2a). In single crystalline materials, however, AMR reflects the crystal symmetry [20, 21]. For centrosymmetric FMs, such as FCC Ni, with $I$ in the $x([100])$, the resistivities with field along the $y[010]$ and $z[001]$ are the same, i.e., for the resistivities with field along the $x$, $y$, and $z$ axes, $\rho_y = \rho_z$ with $\rho_y = \rho_z < \rho_x$. However, the $yz$ scan is *not* constant but shows a 4-fold symmetry. This is observed in a single crystal Ni as shown in Fig. 2a, where there are 4 maxima at $[0\pm10]$ and $[00\pm1]$ separated by 4 minima at $[0, \pm1, \pm1]$.

In contrast, for MR of $Fe_{0.85}Co_{0.15}Si$ (open symbols, Fig. 2b), all three resistivities $\rho_x$, $\rho_y$ and $\rho_z$, are different with $\rho_y < \rho_z < \rho_x$. Equally unusual, all *three xy*, *xz* and *yz* field scans show a *two-fold* symmetry (solid lines, Fig. 3b). Thus, the MR of this B20 magnet reveals the broken $C_4$ symmetry and only $C_2$ prevails. Note that, above $T_C$ (inset of Fig. 2c), MR curves of $Fe_{0.85}Co_{0.15}Si$ for field along $x$, $y$ and $z$ directions are the same within 0.02% accuracy (MR curve for $H_x$ is slightly different from others due to the geometric effect of ordinary magnetoresistance due to the Lorentz force). The magnitude of MR of $Fe_{0.85}Co_{0.15}Si$, as represented by $\rho_z - \rho_y$, decreases with increasing temperature from 2.1 μΩ·cm at 5K and vanishes at and above $T_C$, as shown in Fig. 2c, unequivocally showing the magnetic origin of this unusual MR. The noncentrosymmetric cubic B20 magnets lack the $C_4$ symmetry but only $C_2$ symmetry is observed experimentally (see supplemental materials for more details). This is the first report of the experimental signature revealing directly the broken $C_4$ rotation symmetry.

To theoretically account for the experimental results, one needs a Hamiltonian that preserves the symmetries of $C_2$, $C_3$, and time reversal (*T*), but breaks the inversion (*I*) and $C_4$ symmetries. The microscopic D-M interaction between two neighboring spins $S_i$ and $S_j$ located at $r_i$ and $r_j$ respectively has the form of $D_{ij} \cdot (S_i \times S_j)$. This well-known interaction favors $S_i$ and $S_j$ to be perpendicular to each other and be situated in a plane perpendicular to $D_{ij}$, which is parallel to along the line joining the two spins $|r_i - r_j|$. When an electron of spin $\sigma$ moves from $r_i$ to $r_j$, its spin must rotate as if under a magnetic field along the momentum direction. Thus an electron effectively experiences a magnetic field along its trajectory. This leads to an interaction term of $\alpha \mathbf{k} \cdot \boldsymbol{\sigma}$, where $\mathbf{k}$ is the electron momentum and $\alpha$ measures the strength of this special spin-orbit coupling imposed by the D-M interaction. The ($\mathbf{k} \cdot \boldsymbol{\sigma}$) term clearly preserves the time reversal ($\mathbf{k} \Rightarrow -\mathbf{k}, \boldsymbol{\sigma} \Rightarrow -\boldsymbol{\sigma}$) symmetry but breaks the inversion ($\mathbf{k} \Rightarrow -\mathbf{k}, \boldsymbol{\sigma} \Rightarrow \boldsymbol{\sigma}$) symmetry. It is simple to see that ($\mathbf{k} \cdot \boldsymbol{\sigma}$) also preserve $C_2$ and $C_3$ symmetry, but unfortunately also preserves $C_4$ symmetry. For example, under a $C_4$ rotation about $z$[001], the operation of $(k_x, k_y, k_z) \Rightarrow (k_y, -k_x, k_z)$ and $(\sigma_x, \sigma_y, \sigma_z) \Rightarrow (\sigma_y, -\sigma_x, \sigma_z)$ leaves ($\mathbf{k} \cdot \boldsymbol{\sigma}$) intact, thus unacceptable. In fact, the ($\mathbf{k} \cdot \boldsymbol{\sigma}$) term, with the full rotation symmetry, generates only a constant MR with no directional dependence. Spin-orbital terms with quadratic momentum do break the time reversal symmetry, thus also unacceptable. Cubic terms in momentum are thus required for the unusual transport properties as well as other response properties.

To this end, an effective Hamiltonian for the conduction electrons are constructed based on the theory of invariants [22]. The minimal two-band effective Hamiltonian of the conduction electrons is given by [23]:

$$H = \frac{\hbar^2 k^2}{2m} - \alpha(k_x\sigma_x + k_z\sigma_z + k_z\sigma_z) + \beta[k_x\sigma_x(k_y^2-k_z^2) + k_y\sigma_y(k_z^2-k_x^2) + k_z\sigma_z(k_x^2-k_y^2)] \quad (2)$$

up to cubic orders of momentum $\mathbf{k}$. In addition to the conventional quadratic kinetic energy, two spin-orbital coupling terms are involved, where $\alpha$ and $\beta$ are their coefficients respectively. The inversion symmetry is transparently broken by both spin-orbital coupling terms. The second term in Eq. (2) is dominant around the $\Gamma$ point, and is compatible with the spin-spin interaction in the B20 compounds as mentioned above. Importantly, the third term in Eq. (2) preserves $T$, $C_2$, and $C_3$ symmetries but breaks $I$ and $C_4$ symmetry. Therefore our effective Hamiltonian is a minimal one for the B20 compounds that all essential symmetries are preserved, while any redundant symmetries are absent. Our calculations show that the unusual MR appears *only* when both spin-orbital coupling terms in Eq.(2) are present [23].

The physical picture of the $C_4$ symmetry breaking can be seen by examining the Fermi surface (e.g., $\rho_{yy}$ and $\rho_{xx}$) of the effective Hamiltonian with magnetization $M$ fixed along the $z$ direction. Due to the permutation symmetry, longitudinal resistivities $\rho_{yy}$ and $\rho_{xx}$ are actually $\rho_y$ and $\rho_z$ in experiments respectively. The unconventional MR, namely the difference between $\rho_{xx}$ and $\rho_{yy}$, corresponds to the asymmetry between $k_x$ and $k_y$, which appears only when $\alpha$, $\beta$, and $M$ are all nonzero. A typical Fermi surface at finite $k_z$ is shown in the inset of Fig. 3b, which is elongated along the $x$ direction. Note that once $\alpha = 0$, although the Fermi surface is elongated at finite $k_z$, it relates to the Fermi surface at $-k_z$ by a rotation of $\pi/2$ about $z$ axis. Therefore after averaging over $k_z$, electric transports along $x$ or $y$ directions (that is $\rho_y$ and $\rho_z$ in experiments) are still the same. The effective Hamiltonian in Eq. (2) is the minimal Hamiltonian capturing the observed features.

The conductivity is obtained from the current density correlation calculations, the details of which will be published elsewhere[23]. Here we show only the results relevant to the experiments. As shown in Fig. 3a, the difference between $\rho_z$ and $\rho_y$ appears, and the magnitude of the MR $\rho_z - \rho_y$ decreases with the carrier mean free path $l$. This is consistent with the physical picture we presented that in the large $l$ limit, the carriers are oblivious to the detailed structure within the unit cell and insensitive to the broken $C_4$ symmetry. Thus B20 magnets with high resistivity actually facilitate the observation of the unusual MR due to $C_4$ symmetry breaking. Furthermore, $\rho_z - \rho_y$ increases with the magnetization $M$ (Fig. 3b), which is consistent with experimental results shown in Fig. 3c. Above $T_C$ where magnetization is zero, $\rho_z - \rho_y = 0$. Below $T_C$, the increasing magnetization leads to an increasing $\rho_z - \rho_y$.

In summary, the fascinating properties of the cubic B20 symmetry are due to the broken inversion and broken four-fold ($C_4$) rotation symmetry. We have observed a direct consequence of the broken $C_4$ rotation symmetry via the MR measurements. This observation not only deepens our understanding between symmetry and spin-dependent transport, but also constraints the minimal Hamiltonian consistent with the symmetry. We have proposed an effective Hamiltonian with two spin-orbital coupling terms that satisfies symmetry requirements and reproduces the observed MR results. It has been well known that spin-orbital coupling is the origin of asymmetric spin interactions in B20 chiral magnets. Our result shows that spin-orbital coupling is equally essential for electron transports in these materials. This effective Hamiltonian can be broadly used in future studies of transports in B20 compounds.

We thank Professor Oleg Tchernyshyov for helpful discussions. The work at JHU has been supported by the US NSF DMR-1262253. J. Z. is supported by the Theoretical Interdisciplinary Physics and Astrophysics Center and by the US DOE DEFG02-08ER46544.

Fig.1 (color online). (a) (left) Crystal structure of cubic B20 $Fe_{1-x}Co_xSi$ with same lattice constants $a$, $b$ and $c$; (right) cuboid single crystals ($Fe_{0.85}Co_{0.15}Si$ or Ni) with edges along [100], [010] and [001] directions (defined as $x$, $y$ and $z$ axes), respectively. (b) Magnetic hysteresis loops of $Fe_{0.85}Co_{0.15}Si$ single crystal at 10 K for field along $y$ ([010]) and $z$ ([001]) directions. Inset: ac susceptibility as a function of temperature showing $T_C \approx$ 23 K. (c) Resistivity of $Fe_{0.85}Co_{0.15}Si$ single crystal as a function of field at 3 K for field along $y$ ([010]) and $z$ ([001]) directions. Inset: Resistivity as a function of field above 2 kOe at 5K for field along $x$ and $y$ directions.

Fig.2 (color online). (a) Resistivity at 300 K and $H$ = 10 kOe > $H_S$ as a function of field angle with field in the $yz$ plane for Ni single crystal (open circles) and Ni polycrystal (square symbols). Solid line is $\sin^2 \theta \cos^2 \theta$ fitting to data. (b) Resistivity (linear background subtracted) at 3 K and $H$ = 2 kOe > $H_S$ as a function of field angle for $Fe_{0.85}Co_{0.15}Si$ single crystal with field in the $xy$ plane (open squres), $xz$ plane (open circles) and $yz$ plane (open triangles). Solid lines are $\cos^2 \theta$ fitting to data. (c) Magnitude of MR ($\rho_z$ - $\rho_y$) of $Fe_{0.85}Co_{0.15}Si$ as a function of temperature. Inset: Resistivity as a function of field at 30 K (>$T_C$).

Fig.3 (color online). Theoretically calculated resistivity of $\rho_z$ (open triangles) and $\rho_y$ (open circles) as a function of (a) mean free path and (b) magnetization. Inset in (b): Fermi surface with nonzero $\beta$ and finite $k_z$.

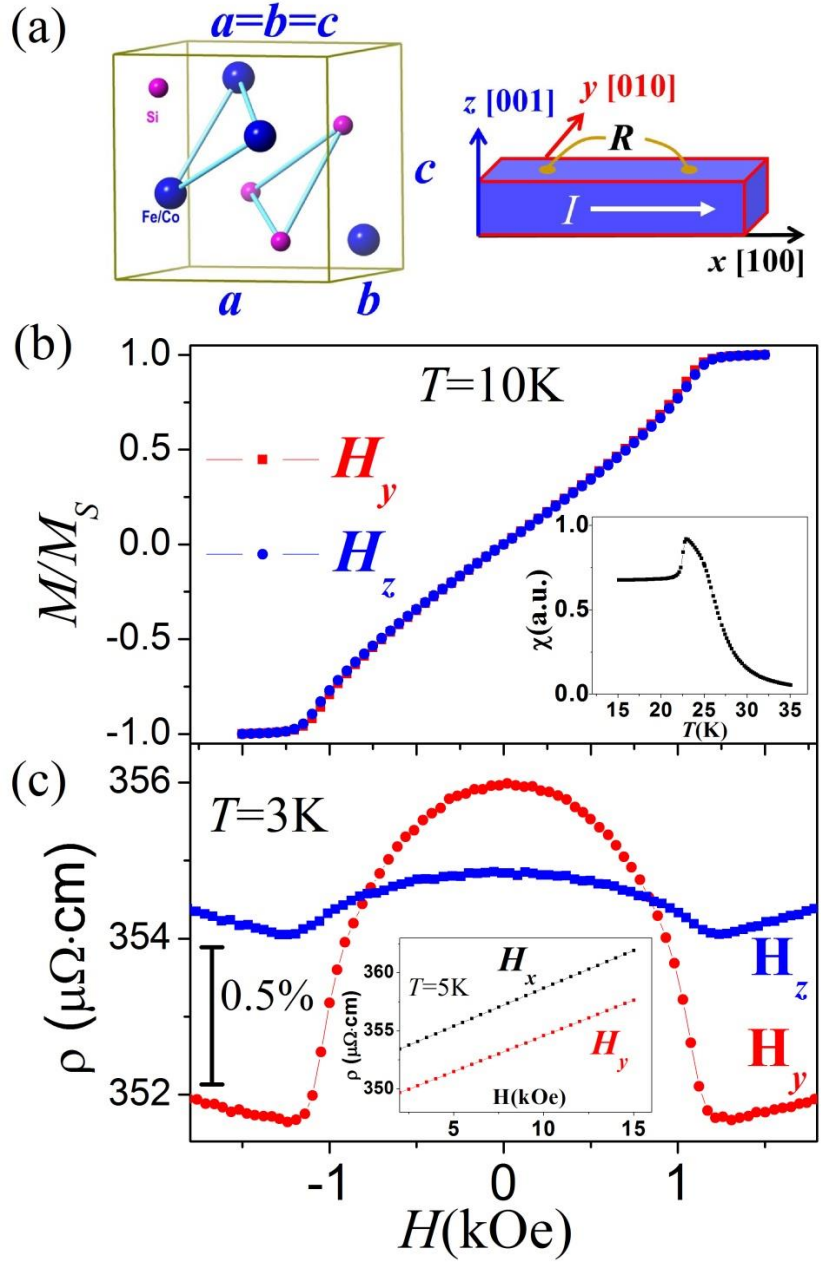

Fig.1 (color online). (a) (left) Crystal structure of cubic B20 $Fe_{1-x}Co_xSi$ with same lattice constants *a*, *b* and *c*; (right) cuboid single crystals ($Fe_{0.85}Co_{0.15}Si$ or Ni) with edges along [100], [010] and [001] directions (defined as *x*, *y* and *z* axes), respectively. (b) Magnetic hysteresis loops of $Fe_{0.85}Co_{0.15}Si$ single crystal at 10 K for field along *y* ([010]) and *z* ([001]) directions. Inset: ac susceptibility as a function of temperature showing $T_C \approx 23$ K. (c) Resistivity of $Fe_{0.85}Co_{0.15}Si$ single crystal as a function of field at 3 K for field along *y* ([010]) and *z* ([001]) directions. Inset: Resistivity as a function of field above 2 kOe at 5K for field along *x* and *y* directions.

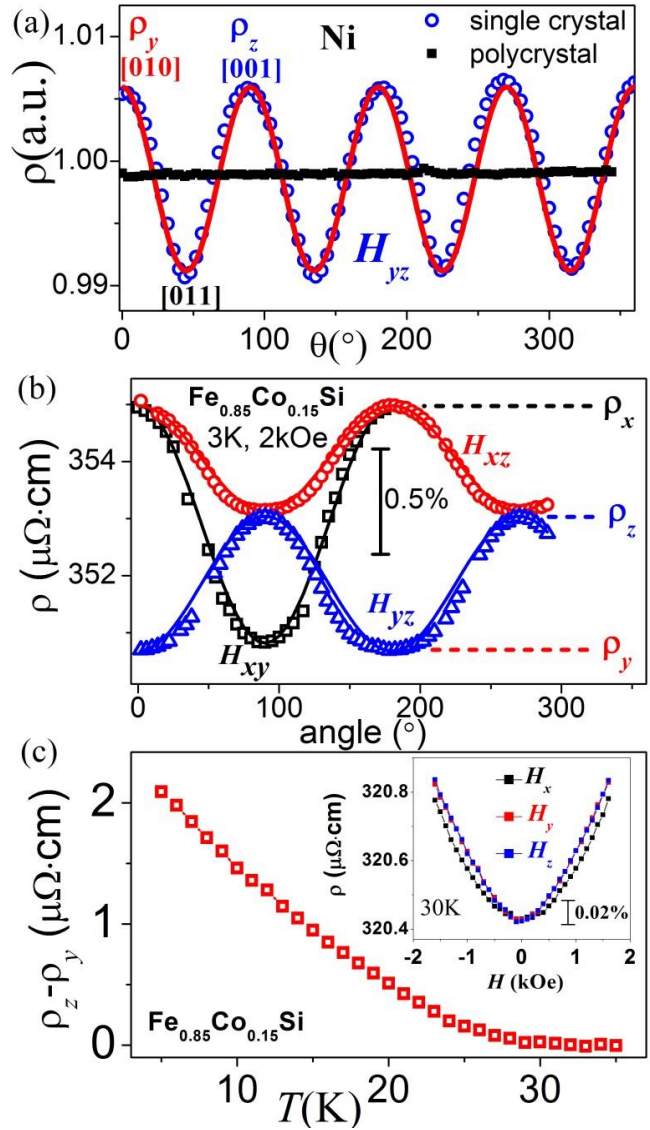

Fig.2 (color online). (a) Resistivity at 300 K and $H = 10$ kOe $> H_S$ as a function of field angle with field in the $yz$ plane for Ni single crystal (open circles) and Ni polycrystal (square symbols). Solid line is $\sin^2\theta\cos^2\theta$ fitting to data. (b) Resistivity (linear background subtracted) at 3 K and $H = 2$ kOe $> H_S$ as a function of field angle for $Fe_{0.85}Co_{0.15}Si$ single crystal with field in the $xy$ plane (open squres), $xz$ plane (open circles) and $yz$ plane (open triangles). Solid lines are $\cos^2\theta$ fitting to data. (c) Magnitude of MR ($\rho_z - \rho_y$) of $Fe_{0.85}Co_{0.15}Si$ as a function of temperature. Inset: Resistivity as a function of field at 30 K ($>T_C$).

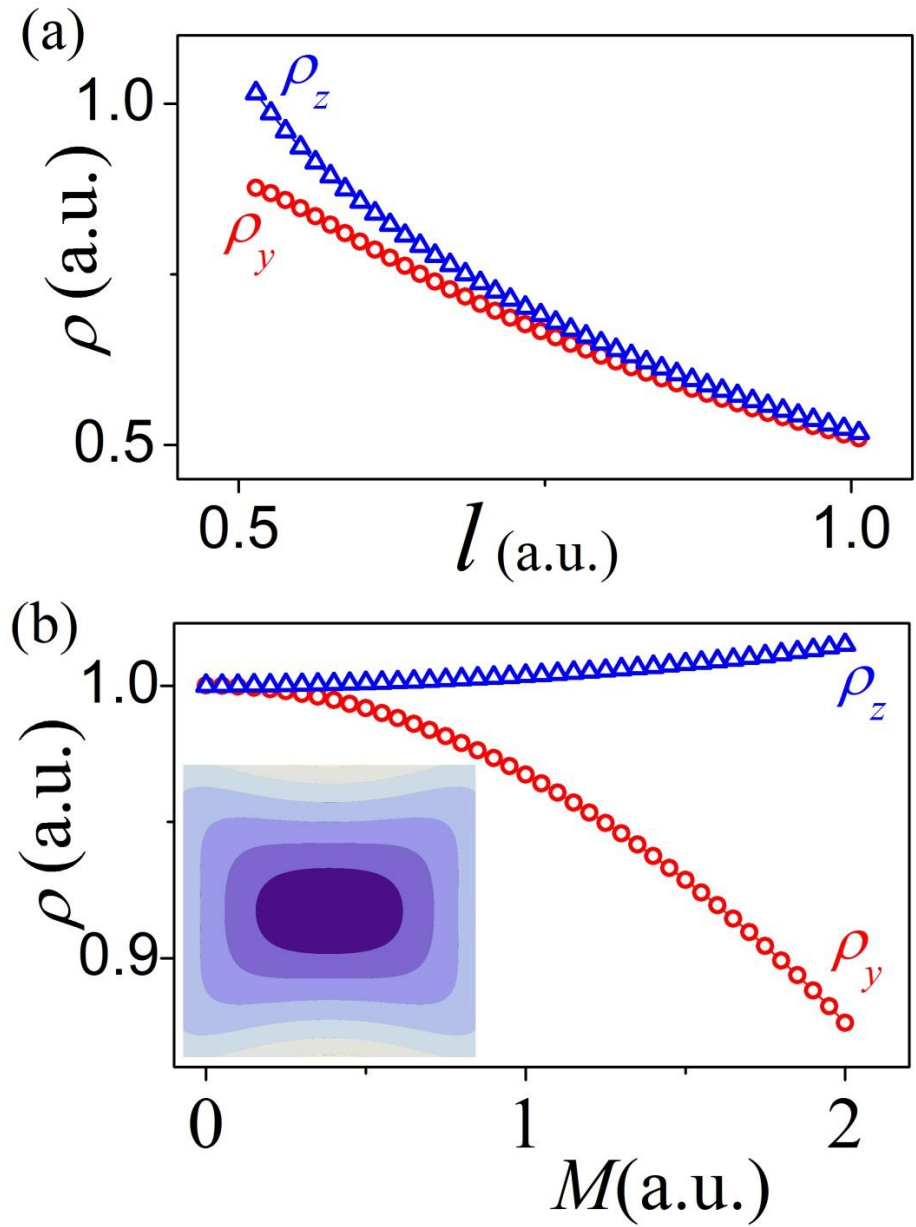

Fig.3 (color online). Theoretically calculated resistivity of $\rho_z$ (open triangles) and $\rho_y$ (open circles) as a function of (a) mean free path and (b) magnetization. Inset in (b): Fermi surface with nonzero $\beta$ and finite $k_z$.

*To whom correspondence should be addressed:

sxhuang@physics.miami.edu

jiadongzang@gmail.com

clc@pha.jhu.edu

#Present address: *Department of Physics, University of Miami, Coral Gables, FL 33146, USA*